\documentclass{moriond}

\usepackage{subfigure}
\newcommand{\invisible}[1]{}

\def \etmis {\mbox{\ensuremath{E\kern-0.6em\slash_T}}}
\def \etmiss {\mbox{\ensuremath{E\kern-0.6em\slash_T}}}

\newcommand{\INVISIBLE}[1]{}

\newcommand{\dzero}  {\ensuremath{\mathrm{D\O}}}

\newcommand{\gev}  {\ensuremath{\mathrm{ GeV}}}

\newcommand{\egev}{\gev}

\newcommand{\pbinv}{\ensuremath{\mathrm{ pb^{-1}}}}

\newcommand{\invfb}{\ensuremath{\mathrm{ fb^{-1}}}}

\def\eg{{\it e.g.}}

\newcommand{\pt}{{\ensuremath{ p_T}}}

\newcommand{\ipb}  {\pbinv}

\newcommand{\BEA}{\begin{eqnarray}}
\newcommand{\EEA}{\end{eqnarray}}

\def\beq  {\begin{equation}}
\def\eeq  {\end{equation}}

\def\PRD#1#2#3{{ Phys. Rev. }{\bf D#1 }(#2) #3}

\newenvironment{packed_itemize}{
\begin{itemize}
  \setlength{\itemsep}{1pt}
  \setlength{\parskip}{0pt}
  \setlength{\parsep}{0pt}
}{\end{itemize}}

\def\PRD{{\em} Phys.\ Rev.\ D}

\def\be{\begin{equation}}
\def\ee{\end{equation}}
\def\bea{\begin{eqnarray}}
\def\eea{\end{eqnarray}}

\newcommand{\Photo}{\includegraphics[height=35mm]{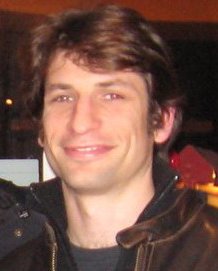}}

\begin{document}
\vspace*{4cm}
\title{HIGGS BOSON PRODUCTION AND PROPERTIES AT THE TEVATRON
}

\author{Boris Tuchming  \\(for the CDF and \dzero\ Collaborations)}

\address{ Irfu/SPP, CEA Saclay,\\ 91191 Gif-Sur-Yvette,  France}

\maketitle\abstracts{
We present the searches for the
Standard Model Higgs boson, using the full Run~II
dataset of the Fermilab Tevatron $p\bar p$ collider,
collected with the  CDF and \dzero\ detectors.
A significant excess of events is observed,
consistent with the presence of a Standard Model Higgs boson of mass 125~\gev.
We also present  tests of different spin/parity hypotheses, performed
in the $VH \to V b\bar b$ channels at \dzero,
and new searches in invisible modes conducted at CDF.}

\section{Introduction }

\label{intro}

Now that the Higgs boson of mass $m_H=125$~\gev\ has been discovered in
2012 by the ATLAS and CMS Collaborations at LHC~\cite{bib:atlas-hdiscovery,bib:cms-hdiscovery},
a new era of measurement has started.
The Run~II of the Tevatron $p\bar p$ collider at $\sqrt s=1.96$~TeV started in 2001 and ended in 2011.
Over a decade, the results of the Tevatron collaborations, CDF and \dzero, 
have been a bridge between the search era and the measurement era.
They provided the first post-LEP constraints on the Standard Model (SM) Higgs boson  mass~\cite{Bernardi:2008ee,bib:tevcomb-hww}, as well as the evidence that Higgs bosons couple to b-quarks~\cite{bib:TeVhbb}.

This proceedings summarizes the combined results from the Tevatron collaborations (see Refs.~\cite{{cdfprd},{d0prd},{CDFDOcombo}} and references therein) using the full Run~II dataset which corresponds to $\sim 10$~\invfb\  of $p\bar p$ collisions per experiment. Recent studies on spin/parity and invisible modes are also presented.

\section{Search channels and strategy}

Within the SM, the branching ratios and the production cross-sections as a function of the Higgs boson mass are well known.
 Over the mass range $90<M_H<200$~\egev, the dominant production process is the gluon-gluon fusion $gg\rightarrow H$ (950~fb  for $M_H=125$~\gev), followed by  the associated production with a weak vector boson $p\bar p\to WH,\ p\bar p\to ZH$ (respectively 130~fb and 79~fb  for $M_H=125$~\gev). The main decay modes   are $H\to b\bar b$  (58\% for $M_H=125$~\gev) and $H\to W^+W^-$ (22\% for $M_H=125$~\gev), so that the most sensitive signatures are:
i) one lepton + \etmis\ + two $b$-jets   (mainly $WH\to \ell\nu b \bar b$),
ii) no lepton + \etmis\ + two $b$-jets   (mainly $ZH\to \nu \bar\nu b \bar b$),
iii) two leptons  + two $b$-jets   ($ZH\to \ell^+\ell^- b\bar b$), and
iv) two leptons + \etmis\  ($H\to W^+W^- \to  \ell^+\nu\ell^-\bar\nu$).
Thus, the Higgs physics at Tevatron mainly relies on
$b$-tagging efficiency, good dijet mass resolution, high-\pt\ lepton acceptance,
good modeling of the \etmis, and good modeling of the $V$+jet background (where $V=W$ or $Z$).
The Tevatron sensitivity to $VH\to V b\bar b$ is complementary to the LHC main discovery channels ($H\to\gamma\gamma,\ H\to ZZ$).

The main sensitivity is given by the four channels presented above, but many other signatures are also considered
to bring additional sensitivity and test the agreement with the SM expectations.
For examples, Tevatron experiments have also looked for diphoton events ($H\to\gamma\gamma$), associated production with top-quark pairs ($t\bar t H$),
lepton + \etmis\ + dijet signatures (from $H\to WW$),
 trilepton signatures (\eg\ from $WH\to WWW$), same charge dilepton signatures (\eg\ from $WH\to WWW$),
quadrilepton signatures (\eg\ from $ZH\to \ell^+\ell^- WW$),
and tau-based signatures (\eg\ from $WH\to q\bar q \tau^+\tau^- $).

Over the course of Run~II,
CDF and \dzero\ have followed the same strategy to optimize the analyses and improve their sensitivity:
\begin{packed_itemize}
\item  Acceptance is maximized
by lowering kinematic requirements on leptons, by including different lepton reconstruction categories,
by accepting events from all possible triggers, and by optimizing object identification with sophisticated multivariate (MVA) techniques (\eg\ $b$-tagging).
\item 
MVA techniques are widely used to maximize use of available information.
Using a MVA as a final discriminant typically provides 25\% more sensitivity
than just using a  single kinematic discriminant such as the dijet mass for the $VH\to Vb\bar b$ channels.
Dedicated MVA are also trained to split analyses into subchannels enhanced or enriched in specific backgrounds. 

\item The various channels are split  into subchannels according to
jet multiplicity, lepton flavor or lepton quality, and $b$-tagging content.
Using subchannels with different signal-over-background ratio ($s/b$) maximizes discriminating power, allows sensitivity to different signal production modes, and provides more handles and lever-arm to control backgrounds and systematic uncertainties.

\item
The data are employed  as much as possible. Instrumental backgrounds, such as jets or photons faking leptons, charge mismeasurements,
and tail of \etmis\ resolution are measured in dedicated control samples.
Background enriched samples are also employed to check modeling of specific background processes.
Eventually, the same analysis techniques, namely the same kind of MVA,
the same subchannels, and the same treatment of systematic uncertainties
are employed to measure production rates of known SM candles such as
$p\bar p\to W^+W^-\to \ell^+\nu\ell^-\bar\nu$, or $VZ \to V b \bar b$.
For example,
the combined CDF+\dzero\ measured cross section $\sigma(WW+WZ)=3.0\pm 0.6\ \rm{stat} \pm 0.7\ \rm{syst}~\ipb$ is in agreement with SM prediction of $4.4\pm0.3~\ipb$.

\end{packed_itemize}

\section{Higgs boson studies}

\subsection{Search for Standard Model Higgs boson}
\begin{figure}[!htb]
\centering
\subfigure[Background p-value as a function of the Higgs boson mass hypothesis.\label{fig:Clb}]{\includegraphics[width=0.48\textwidth]{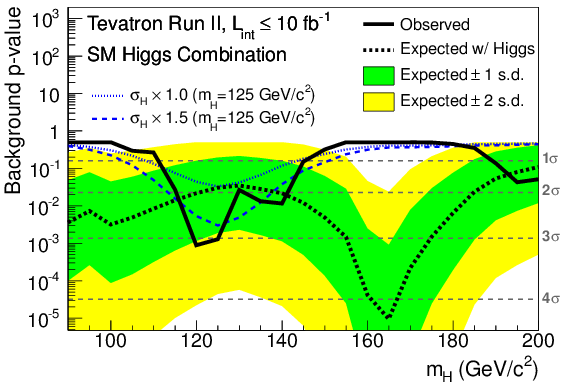}}
\hfill\strut
\subfigure[Limits on the SM Higgs boson production  as a function of the Higgs boson mass hypothesis.\label{fig:tevlimit}]
{\includegraphics[width=0.48\textwidth]{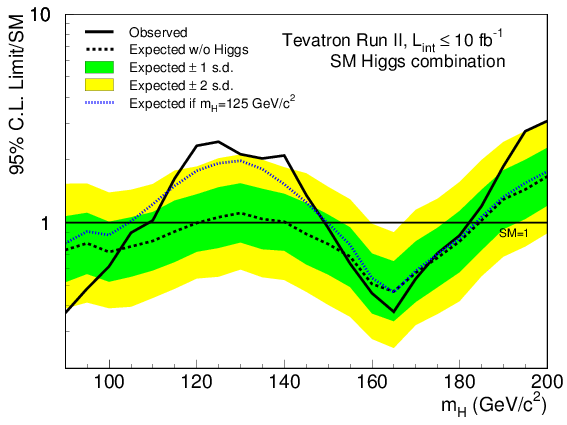}}
\caption{CDF and \dzero\ combined results.}
\end{figure}

The results from the different search channels are combined
with a log-likelihood ratio (LLR)
testing the signal-plus-background over the background-only hypothesis as a function of the Higgs boson mass.
The background p-value arising from this test is shown
in Fig.~\ref{fig:Clb}.
A significant signal-like excess in the mass range between 115 and 140~GeV is observed.
The background  p-value of that excess corresponds to 3.0 standard deviation (s.d.) for $M_H=125~\gev$.
That excess arises from both CDF (2.0 s.d.) and \dzero\ (1.7 s.d) data.
The LLR test statistic is also employed to
derive limits at 95\% C.L.\ on the Higgs boson production measured in units of the expected SM production. The limits
are shown in Fig.~\ref{fig:tevlimit}.
The combined CDF and \dzero\ results almost reaches the exclusion sensitivity over the full range $[90,185]~\gev$.
Because of the excess observed in the low mass region, the actual observed exclusion ranges are: $[90,109]~\gev$ and  $[149,182]~\gev$, which is smaller than expected.

\subsection{Measurement of production rates}
The SM search channels  can be separately combined to measure
the yield in the different modes: $H\to b\bar b$, $H\to\tau^+\tau^-$, $H\to W^+W^-$, and $H\to \gamma\gamma$. The best fits to the data are summarized in Table~\ref{tab:SM_yield}.
Relative to the SM expectation, the overall production rate 
$R=1.44^{+0.59}_{-0.56}$  is measured for $M_H=125~\gev$.
The modes with sizable signal-like excesses relative to the background-only hypothesis are  $VH\rightarrow Vb{\bar{b}}$  and $H\rightarrow W^+W^-$, as expected from the SM Higgs boson.
The most sensitive channel is  $VH\rightarrow Vb{\bar{b}}$ with a fitted production rate of $R=1.6\pm0.7$.
This result is competitive with respect to the measurements of
$R(VH\rightarrow Vb{\bar{b}})=0.2\pm0.6$ and 
$R(VH\rightarrow Vb{\bar{b}})=1.0\pm0.5$ obtained respectively by
the ATLAS~\cite{Atlas:Vbb} and CMS~\cite{Chatrchyan:2013zna} Collaborations.

\begin{table}[!b]\center
\begin{tabular}{|l|c|c|c|}\hline\rule[-2mm]{0mm}{6mm}
   &  CDF~\cite{cdfprd}  & \dzero~\cite{d0prd}  &  CDF+\dzero~\cite{CDFDOcombo}\\ \hline
\rule[-2mm]{0mm}{7mm}
$R_{\rm{fit}}$($H\rightarrow W^+W^-$)      & $0.00^{+1.78}_{-0.00}$      & $1.90^{+1.63}_{-1.52}$        & $0.94^{+0.85}_{-0.83}$  \\
\rule[-2mm]{0mm}{7mm}
$R_{\rm{fit}}$($VH\rightarrow Vb{\bar{b}}$) &$1.72^{+0.92}_{-0.87}$         & $1.23^{+1.24}_{-1.17}$      & $1.59^{+0.69}_{-0.72}$ \\
\rule[-2mm]{0mm}{7mm}
$R_{\rm{fit}}$($H\rightarrow \gamma\gamma$)&$7.81^{+4.61}_{-4.42}$       &  $4.20^{+4.60}_{-4.20}$     & $5.97^{+3.39}_{-3.12}$ \\ 
\rule[-2mm]{0mm}{7mm}
$R_{\rm{fit}}$($H\rightarrow \tau^+\tau^-$)&$0.00^{+8.44}_{-0.00}$        &         $3.96^{+4.11}_{-3.38}$ & $1.68^{+2.28}_{-1.68}$  \\
\rule[-2mm]{0mm}{7mm}
$R_{\rm{fit}}$($t\bar tH\rightarrow t\bar t b\bar b$)&$9.49^{+6.60}_{-6.28}$        &  --                       & --  \\\hline
\rule[-2mm]{0mm}{7mm}
$R_{\rm{fit}}$(combined SM)                       &  $1.54^{+0.77}_{-0.73}$        &   $1.40^{+0.92}_{-0.88}$     & $1.44^{+0.59}_{-0.56}$ \\ \hline

\end{tabular}
\caption{
\label{tab:SM_yield}
Best fit to the data of the Higgs boson production (in unit of the SM Higgs boson production), assuming $M_H=125~\gev$,
for the different channels and their combination.
}
\end{table}

\subsection{Measurement of couplings to fermions and bosons}

Assuming a SM-like Higgs particle of 125~\gev,
the SM couplings  to fermions and vector bosons are scaled by respectively
$\kappa_f$, $\kappa_W$, and $\kappa_Z$, accounting also for the overall scaling of the total width.
A fit to the data is performed by scaling properly the contributions from the different production and decay modes, and 2-dimension and 1-dimension posterior density probability are obtained for the coupling scale factors.
The 1-dimension constraints on the coupling scale factors are: 
i) assuming $\kappa_W=\kappa_f=1$, the best-fit value is $\kappa_Z=\pm1.05^{+0.45}_{-0.55}$;
ii) assuming $\kappa_Z=\kappa_f=1$,  the best-fit 68\%\ confidence intervals
are defined by  $\kappa_W=- 1.27^{+0.46}_{-0.29}$ and      $1.04<\kappa_W <1.51$;
iii) assuming $\kappa_W=\kappa_Z=1$, the best-fit value
is $\kappa_f=-2.64^{+1.59}_{-1.30}$;
iv) and by letting $\kappa_f$ floating with a flat prior, 
the custodial symmetry is tested and the best fit value for the ratio
 $\lambda_{WZ}=\frac{\kappa_W}{\kappa_Z}$ reads  $\lambda_{WZ}=1.24^{+2.34}_{-0.42}$.
All these results are in agreement with the SM expectations within their uncertainties.

\subsection{Spin and parity tests}
The tests are based on the property that  spin and parity of a particle affects the shape of the
excitation curve near the production threshold. Thus,
the spectra of the effective center-of mass energy, $\sqrt{\hat s}$,
of $VH\to V b\bar b$ events are expected to be quite different under different
spin and parity hypotheses ($0^-$, $0^+$, or $2^+$) for $H$~\cite{bib:Elis}.
This property  is exploited
by \dzero\ to re-analyze the data samples from the $VH\to V b\bar b$ 
SM Higgs channels~\cite{d0conf:JP}. \dzero\ uses as discriminant observable the 
overall mass (or transverse mass for final states with neutrinos) of the candidate events.
The signal sensitivity is enhanced by splitting the samples into low and high
purity regions, according to the dijet invariant mass ($ZH$ channels)
or the  SM MVA  discriminant output ($WH$ channel).
The results are obtained assuming the production times branching fraction ($\sigma\times Br$) of  1.23$\times $SM (\dzero\ best fit value for $H\to b\bar b$):  i)
a $0^+$ hypothesis is favored over a $0^-$ and $2^+$  signal at the
99.9\% and 99.5\%, respectively; ii) a mixture of $0^-$ and $0^+$ signals is excluded at 95\% C.L.\ for fractions of $0^-$ signal higher than 0.67; iii)  
 a mixture of $2^+$ and $0^+$ signals is excluded at 95\% C.L.\ for fractions of $2^+$ signal higher than 0.57.

\subsection{Search for invisible decays}
CDF  exploits the $Z\to \ell^+\ell^-$ + large \etmis\ signature to search for $ZH$ where $H$ decays to undetected products~\cite{cdfconf:Hinvisible}.
The absence of excess in the data allows to set a limit of $\sigma\times Br> 90$~fb for invisible modes
for $m_H=125~\gev$. A 100\% branching fraction to invisible particles is excluded if $M_H<120$~\gev.

\section{Conclusion}

The final combined results of CDF and \dzero\ exhibits a 3.0 s.d.\ evidence
for the production of the SM Higgs boson.
Within their respective uncertainties, 
the measurements of production rates and couplings show 
good agreement with the SM expectations.
Recent analyses of spin and parity properties and recent searches for invisible decays also exhibit
consistency with the SM.

\end{document}